\documentclass[aps,prl,twocolumn,showpacs,preprintnumbers,amsmath,amssymb, superscriptaddress]{revtex4}

\usepackage{graphicx}
\usepackage{dcolumn}
\usepackage{bm}

\begin{document}

\preprint{APS/123-QED}

\title
{
Determining the optical axes of entangled Laguerre Gauss modes
}

\author{Daisuke Kawase}
\affiliation{%
Research Institute for Electronic Science, Hokkaido University,
Sapporo 060--0812, Japan
}%

\author{Shigeki Takeuchi}
\altaffiliation[Correspondence address: ]{takeuchi@es.hokudai.ac.jp}
\affiliation{%
Research Institute for Electronic Science, Hokkaido University,
Sapporo 060--0812, Japan
}%
\author{Keiji Sasaki}
\affiliation{%
Research Institute for Electronic Science, Hokkaido University,
Sapporo 060--0812, Japan
}%

\author{Atsushi Wada}
\affiliation{%
Department of Information and Communication Engineering, the University of Electro-Communications
}%
\author{Yoko Miyamoto}
\affiliation{%
Department of Information and Communication Engineering, the University of Electro-Communications
}%

\author{Mitsuo Takeda}
\affiliation{%
Department of Information and Communication Engineering, the University of Electro-Communications
}%

\date{\today}
%

\begin{abstract}
A method for determining the positions of hologram dislocations relative to the optical axes of entangled Laguerre Gaussian modes is proposed. In our method, the coincidence count rate distribution was obtained by scanning the position of one of the holograms in two dimensions. Then, the relative position of the hologram dislocation was determined quantitatively from the positions of the minimum and maximum coincidence count rates. The validity of the method was experimentally verified, and in addition, an experiment demonstrating the violation of the Clauser-Horne-Shimony-Holt inequality was performed using the well-identified optical axes of the entangled modes.
\end{abstract}

\maketitle


\section{I. Introduction}

The quantum states of photons in Laguerre-Gaussian (LG) modes have been attracting a lot of attention recently. It is well known that photons in LG modes have orbital angular momentum, just as circularly polarized photons have spin angular momentum. However, there is a big difference between these two physical properties. In principle, an $N \times N$ Hilbert space with arbitrary large $N$ can be created from a single photon in an LG mode, which is much larger than the $2 \times 2$ Hilbert space obtainable with polarization. Such quantum states, called quNits, can be used to improve quantum information protocols \cite{Kaszlikowski-PRL2000,Cerf-PRL2002,Nihira-PRA2005}. It is also interesting to study the nature of entanglement in larger Hilbert spaces using single photons in LG modes.

In a pioneering paper \cite{Vaziri-Nature2001}, Vaziri et. al. confirmed experimentally that the angular momentum of photon pairs are conserved through the parametric down conversion process and that the generated pairs can be represented as a superposition of states in different LG modes. As a logical conclusion, they claimed that the generated pairs are entangled in LG modes. Later, the same authors reported a violation of Clauser-Horne-Shimony-Holt (CHSH) inequality in three dimensions \cite{Vaziri-PRL2002}. Recently, Langford et. al. performed quantum state tomography with their entangled photonic qutrits and qubits and confirmed entanglement quantitatively \cite{Langford-PRL2004}. Concentration of entanglement in LG modes \cite{Vaziri-PRL2003}, and production, transmission, and reconstruction of qutrits \cite{Molina-Terriza-PRL2004} have also been demonstrated using entangled photons in LG modes.

 In those experiments, holograms with a dislocation in the center were used as LG mode converters \cite{Arlt-PhDthesis}. When the optical axis of the light beam is `exactly' located on the dislocation, a state in a LG mode is converted to one in a Gaussian mode, and is detected by a photon counter through a single mode optical fiber. In contrast, a superposition state of different LG modes is converted to a Gaussian mode when the optical axis is located at an appropriate position relative to the dislocation. Thus, it is crucial to be able to determine the position of the dislocation relative to the optical axis. However, it is not a trivial matter for the case of entangled LG modes, typically generated via spontaneous parametric down conversion (SPDC), because the optical axis of the signal beam have to correspond with the optical axis of the idler beam in accordance with the phase matching condition.

 In this paper, we present a method for determining the position of the hologram dislocation relative to the optical axis of entangled LG modes. In our method, we obtain the coincidence count rate distribution by scanning the position of the hologram relative to the idler mode position in two dimensions. The relative position of the hologram dislocation in the signal beam can then be determined quantitatively from the location of the minimum and maximum points of the coincidence counts\footnote{In ref \cite{Vaziri-PRL2003}, in order to demonstrate the incoming state is entangled in LG modes, position of holograms in one beam were displaced while in the other beam the corresponding holograms performed an one-dimensional (x) scan. In the literature, the high visibility of coincidence count rates was used as a signature of entanglement, however, neither the relative position of the hologram dislocation to the optical axis nor beam waist size have not been analyzed using the measurement result.}.
 We also found that the obtained two-dimensional maps of the coincidence count rates exhibited intuitively a distinctive feature of entanglement, namely, correlations in different non-orthogonal basis sets. From the experimentally obtained two coincidence distributions, we succeeded in showing, using the well-defined optical axes, that violation of the CHSH inequality is a signature of entanglement in LG modes.

 This paper is organized as follows. In section II, we explain how LG modes are converted by reflection-type hologram gratings. In section III, we first explain how we can identify the measurement basis states by considering the time-reversal process, and then we describe the proposed method for determining the position of the optical axis relative to the hologram dislocation. In section IV, we describe our experimental setup. In section V, the procedure used for determining the relative position of the hologram dislocation to the optical axis from experimental measurements is described in detail. In section VI, we report the violation of the CHSH inequality, using the well-defined optical axes of entangled LG modes. Then, we conclude this paper in section VII.


\section{II. Converting LG modes with holograms}

When a pump light in LG$_{0}$ mode is used, the quantum state of photon pairs generated through the process of SPDC is written as follows \cite{Walborn},
\begin{equation} \label{eq:entangle}
|\Phi\rangle = \displaystyle \sum _{m=-\infty}^{+\infty}  \alpha_{m}|-m\rangle_{\textrm{A}}|m \rangle_{\textrm{B}}.
\end{equation} 
Where $|m \rangle$ is a one-photon state of a LG$_{m}$ mode, and $\alpha_{m}$ is the probability amplitude. Here we neglected the radial index of the LG modes since the radial index was not distinguished in our experiment.

\begin{figure}[tbp]
\begin{center}
\scalebox{0.2}[0.2]{\includegraphics{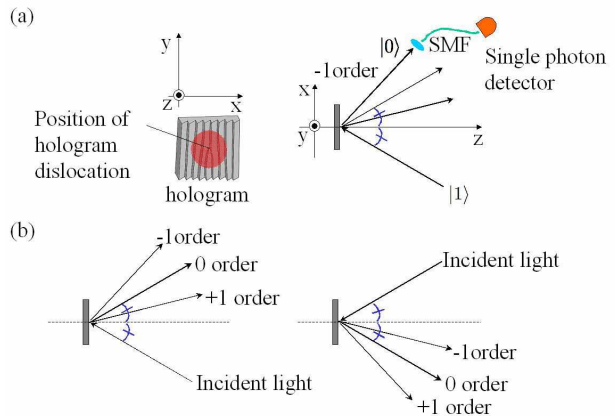}}\\
\end{center}
\caption{The experimental set up for detecting photons in LG$_{1}$ mode. If the dislocation of the hologram is located on the central axis of the incident beam of LG$_{1}$ mode, then the -1st order diffracted beam will be in LG$_{0}$ mode, which couples into the SMF.}
\label{fig:fig1}
\end{figure}

In order to detect photons in a specific superposition of states of LG modes, an optical setup with a hologram and a single mode fiber (SMF) \cite{Superposition} shown in Fig.\ref{fig:fig1}(a) was used. 
Here we assume that a reflection type hologram is used and that it has a one-pronged dislocation at its center \cite{Miyamoto}. We also adopt the definition of diffraction order number given in Fig. \ref{fig:fig1}(b). When the central axis of the incident beam in a LG$_{n}$ mode is centered on the dislocation of the hologram, the m-th order diffracted mode is converted to a LG$_{-n-m}$ mode. Thus, a photon in the LG$_{1}$ mode in the incident beam will be converted to a LG$_{0}$ mode in the $-1$st order diffracted beam. In other words, the photons which are not in the LG$_{1}$ mode will not be converted to LG$_{0}$ mode. The SMF, which is positioned in the $-1$st order diffracted beam, functions as a filter that blocks all the LG modes except for the LG$_{0}$ mode. Thus, by considering the time-reversal of this process, we can conclude that the photons which are detected after the SMF are those which were in the LG$_{1}$ mode in the incident beam.

\begin{figure}[tbp]
\begin{center}
\scalebox{0.22}[0.22]{\includegraphics{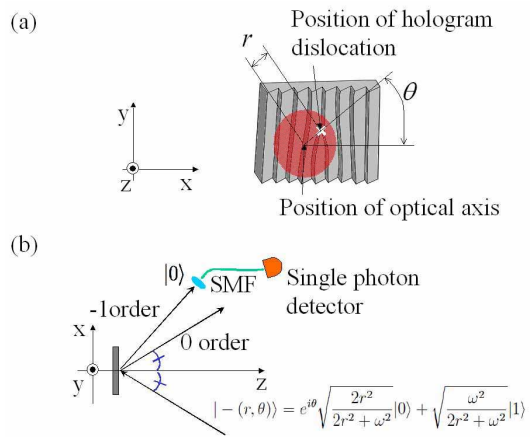}}\\
\end{center}
\caption{An experimental set up for detecting photons in a superposition of LG$_{0}$ and LG$_{1}$ modes. When the dislocation of the hologram is located at the position of the phase singularity of mode, the -1st order diffracted beam becomes the LG$_{0}$ mode which couples into the SMF.}
\label{fig:fig2}
\end{figure}

In order to measure the photons in a superposition state of different LG modes, the position of the dislocation needs to be shifted from the central axis of the incident mode. We define $r$ as the distance between the dislocation in the hologram and the central axis of the optical beam, and $\theta$ as the azimuthal angle illustrated in Fig. \ref{fig:fig2}. Suppose that photons in a Gaussian mode are incident onto the hologram. Since the hologram is designed so that the position of the phase singularity of the diffracted beam is located on the dislocation in the hologram, the diffracted photon will have a superposition of states in the LG$_{0}$ and LG$_{1}$ modes, given, to a good approximation, by the following expression \footnote{Here, the radial index of LG$_{0}$ and LG$_{1}$ modes are implicitly assumed to be 0. Similar equation appears in \cite{Superposition}. Please also refer our comment in section VII.},

\begin{equation} \label{eq: superposition}
e^{(\theta + \pi)i}\sqrt{\frac{2r^{2}}{2r^{2}+\omega^{2}}}|0\rangle+\sqrt{\frac{\omega^{2}}{2r^{2}+\omega^{2}}}| 1\rangle
\end{equation}
, where $\omega$ is the beam radius.
Note that the diffracted photons will also be in other modes (i.e. LG$_{-1}$, LG$_2$, and so on), however, here we neglect those terms since the amplitudes of those states are small
\footnote{Our numerical analysis suggests that the contribution of these additional modes is less than 10.3 \%.}.

When an incident photon is in a pure state of one or two LG modes, it is easy to determine the position of the dislocation in the hologram relative to the optical axis of the modes. For instance, the intensity distribution of the diffracted beam directly gives the position of the phase singularity point, which is where the intensity of the beam is zero. However, this is not the case for entangled photon pairs in LG modes. This is because the quantum state of one photon in the pair is not a pure state, but a mixed state of the various LG modes, and the intensity distribution (the spatial distribution of the single count rates) does not show any singularity.

In the next section, we propose a method for estimating the position of the hologram dislocation relative to the optical axis of the entangled photon pair, which involves taking coincidence count probabilities by scanning the hologram position in two dimensions. 


\section{III. Determining the optical axis position using coincidence count rates - theory }

\begin{figure}[tbp]
\begin{center}
\scalebox{0.2}[0.2]{\includegraphics{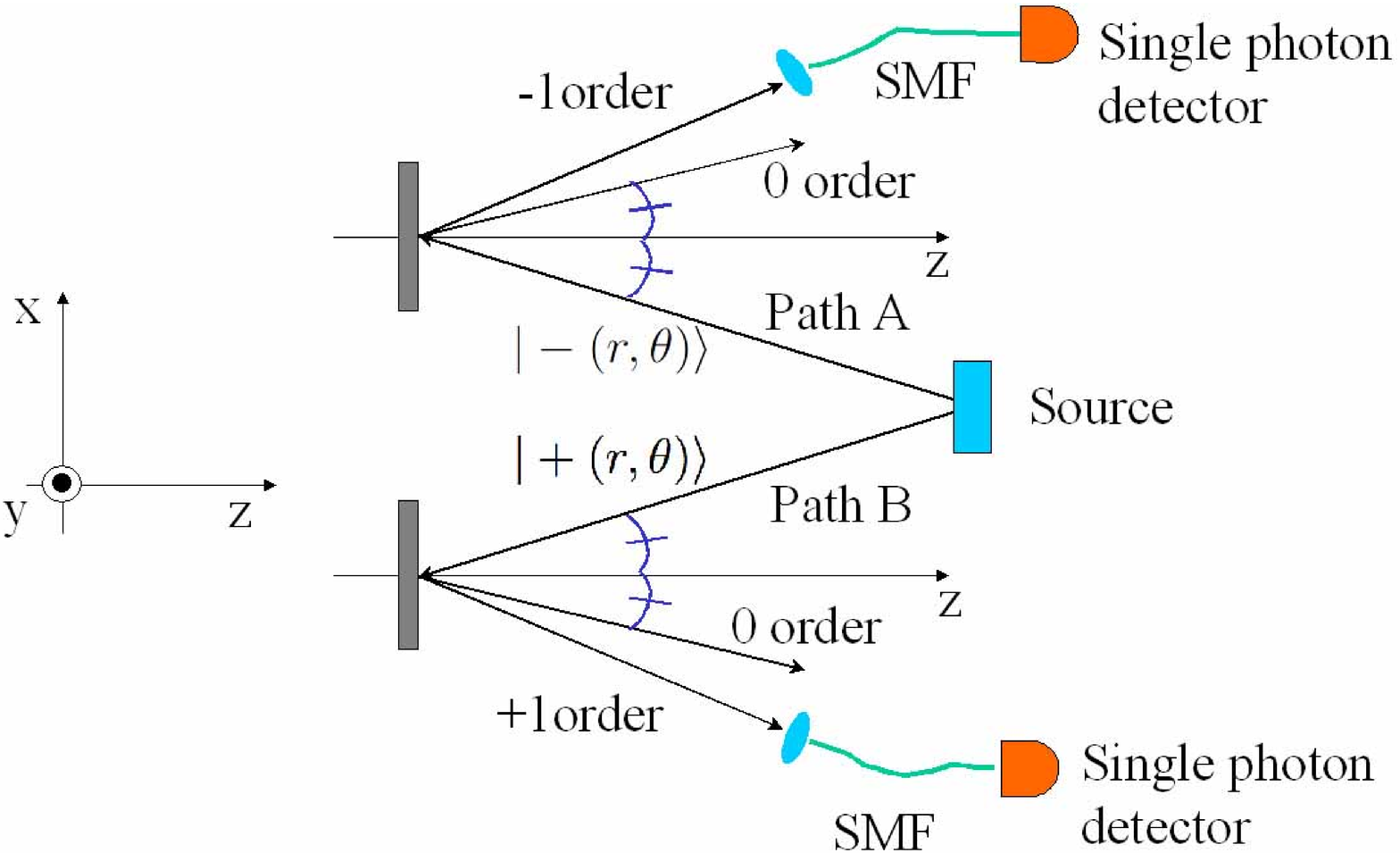}}\\
\end{center}
\caption{An experimental set up for the measurement of correlations between LG modes in path A and path B.}
\label{fig:fig3}
\end{figure}

In the following calculation, we assume that the signal photons and idler photons propagate along the optical paths A and B, and then the -1st order and 1st order diffraction beams are coupled to two SMFs, respectively (Fig. \ref{fig:fig3}). The position of the dislocation relative to the optical axis for each path are $(r_{A},\theta_{A})$ and $(r_{B},\theta_{B})$.

In order to calculate the coincidence count probabilities, we first need to know the measurement basis states for the given optical set up. In other words, we have to identify the `state', which is converted into a Gaussian mode as a result of diffraction by the hologram. This can be found by considering the time-reversal process discussed in the previous section. 

We have already stated that a Gaussian state is converted into the state given by eq.(2). Noting that  $\exp(-i\theta)|m \rangle$ is converted into $\exp(i\theta)(-1)^m|m \rangle$ by the time-inversion operation, we find that the superposition state given by,

\begin{equation} \label{eq:base1}
|-(r,\theta)\rangle = e^{-i\theta}\sqrt{\frac{2r^{2}}{2r^{2}+\omega^{2}}}|0\rangle+\sqrt{\frac{\omega^{2}}{2r^{2}+\omega^{2}}}| 1\rangle ,
\end{equation}
is converted to the Gaussian state $|0\rangle$ in the -1st order diffraction beam. Therefore, eq.(\ref{eq:base1}) is a measurement basis state for path A in our experimental setup.

Using a similar argument, we also find that

\begin{equation} \label{eq:base2}
|+(r,\theta)\rangle = e^{i\theta}\sqrt{\frac{2r^{2}}{2r^{2}+\omega^{2}}}|0\rangle+\sqrt{\frac{\omega^{2}}{2r^{2}+\omega^{2}}}| -1\rangle
\end{equation}
is converted to the Gaussian state $|0\rangle$ in the 1st order diffraction beam. Similarly, eq.(\ref{eq:base2}) is a measurement basis state for path B in our experimental setup.

From eq.(\ref{eq:entangle}), (\ref{eq:base1}), and (\ref{eq:base2}), the coincidence counts probability $P$ for the entangled photon pair is given by
\begin{eqnarray} \label{eq:cc}
& &P_{r_{B},\theta_{B}}(r_{A},\theta_{A})\nonumber\\
&=& |_{A}\langle-(r_{A},\theta_{A})|_{B}\langle+(r_{B},\theta_{B})|\Phi\rangle|^{2}\nonumber\\
&=&|\alpha|^{2}\frac{4r_{A}^{2}r_{B}^{2}+4r_{A}r_{B}\omega^{2} \cos(-\theta_{A}+\theta_{B}+\delta)+\omega^{4}}{(2r_{A}^{2}+\omega^{2})(2r_{B}^{2}+\omega^{2})}
\end{eqnarray}
, where we assume that $\alpha_{-1}=\alpha$ and $\alpha_{0}=\alpha e^{i\delta}$.

This equation suggests that, as a result of entanglement, the coincidence probability has a maximum and a minimum (zero) at particular relative positions of the hologram in path A, for a given relative position $(r_{B},\theta_{B})$ of the hologram in path B. Actually, $P$ has a maximum value $|\alpha|^{2}$ when $(r_{A},\theta_{A})=(r_{B},\delta + \theta_{B})$, and a minimum value 0 when $(r_{A},\theta_{A})=(\omega^{2}/(2r_{B}),\pi+\delta + \theta_{B})$. Such minimum and maximum positions can be found by taking coincidence count probabilities while scanning the position of hologram A. Note that the maximum and minimum positions are always opposite to each other relative to the position of optical axis ($r_A = 0$) between them. Consequently, the distance $d$ between the maximum position and the minimum position is given by
\begin{equation}\label{eq:min-max}
d=r_B+\frac{\omega^2}{2 r_B} .
\end{equation}

 In order to check the validity of this method, we performed the experiment described in the next section.


\section{IV. Experimental setup}
\begin{figure}[tbp]
\begin{center}
\scalebox{0.22}[0.22]{\includegraphics{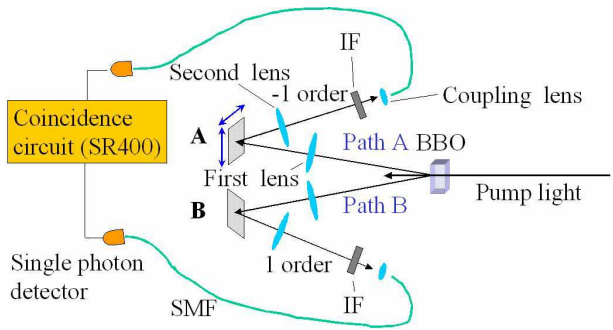}}\\
\end{center}
\caption{Experimental setup used to determine the hologram position}
\label{fig:fig4}
\end{figure}

The experimental set up used is shown in Fig. \ref{fig:fig4}. We used a  $\beta$-barium borate (BBO) crystal cut for Type I phase matching condition having dimensions 5(height) $\times$ 5(width) $\times$ 3(thickness) mm. It was pumped by an argon-ion laser, having a wavelength of 351 nm, a beam radius of 1.2 mm and a power of 150 mW.
The generated signal and idler photons were focused by a plano-convex lens ($f=300$ nm) and the hologram was positioned in the focal plane. The hologram plate was made by fabricating the structure on
thin polymer layer on glass substrate using an electron beam
writer, and then coating with gold. The hologram was 2 mm in diameter. The -1-th (1-th) order diffraction beam in path A (B) was collimated by a plano-convex lens ($f=200$ nm) and filtered by a narrow band-pass filter (IF) having a center wavelength of 702 nm and a FWHM bandwidth of 4 nm. The photons were then coupled to SMFs via objective lenses, and counted by single photon detectors (AQR-FC, Perkin Elmer). The single counting rates and coincidence counting rates were measured by a photon counter (SR-400, Stanford Research System). 

The single counting rates for both paths A and B were maximized by adjusting the positions of the respective coupling lenses. While keeping the position of the hologram B fixed, we measured the coincidence counting rates while scanning in two dimensions the position of the hologram A, which was positioned in the focal plane of the first convex lens.  Typical single counting and coincidence counting rates were 5000 [cps] and 100 [cps], respectively. The vertical and horizontal positions were shifted by a step of 150 $\mu$m for a grid made up of 14 $\times$ 14 points. It took 40 minutes to obtain a complete map of the coincidence count rates. 


\section{V. Determining the optical axis position using coincidence count rates - experiment}


\begin{figure}[tbp]
\begin{center}
\scalebox{0.25}[0.25]{\includegraphics{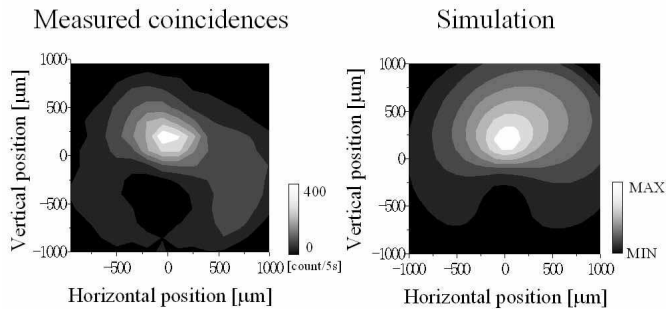}}\\
\end{center}
\caption{A comparison of measurement data with calculation data}
\label{fig:fig5}
\end{figure}

Fig. \ref{fig:fig5} (a) is a contour plot of the coincidence count rates obtained by scanning the position of hologram A for a certain position of hologram B. The horizontal and vertical axes represent the horizontal and vertical position of hologram A relative to the optical axis, respectively. The white area indicates the location of the hologram where the coincidence rates were highest. Note that the measured coincidence counting rates were affected by the position-dependent diffraction efficiency of the hologram we used; the measured efficiency of hologram A decreased as the distance from the dislocation increased. The coincidence rate was almost zero when $r_A$ is large ( $>$ 1mm ). This may be because the incident beam was outside of the hologram area  (2 mm in diameter).

The measured coincidence count distribution had a maximum at $(x,y)=(-50 \mu$m, $200 \mu$m) and a minimum at $(x,y)=(-50 \mu$m, $-400 \mu$m). As discussed in the previous section, the position of the optical axis in path A should be located half-way between these two points. Therefore, $\theta_{B}+\delta$ is estimated to be $\pi/2$. 


In order to estimate the other parameters, $r_{B}$ and $\omega$, we tried to simulate the coincidence probabilities using eq.(\ref{eq:cc}). The position-dependent diffraction efficiency of the hologram, which was experimentally measured using a HeNe laser beam (632 nm) while scanning the focus position, was multiplied by $P(r_A,\theta_A)$ in eq.(\ref{eq:cc}). In Fig. \ref{fig:fig5}(a), the distance between the maximum point and the minimum point was 600 $\mu$m. Thus from eq. (\ref{eq:min-max}), $r_B+\omega^2 / (2 r_B) = 600 \mu$m. We performed several calculations for different $\omega$ and $r_B$ which satisfy this relation and found that the distribution of the high coincidence region is well approximated by $\omega=400 \mu$m and $r_B = 200 \mu$m. The result of simulation is shown in Fig. \ref{fig:fig5}(b). The estimated $\omega$ is consistent with our optical setup. 

\begin{figure}[tbp]
\begin{center}
\scalebox{0.2}[0.2]{\includegraphics{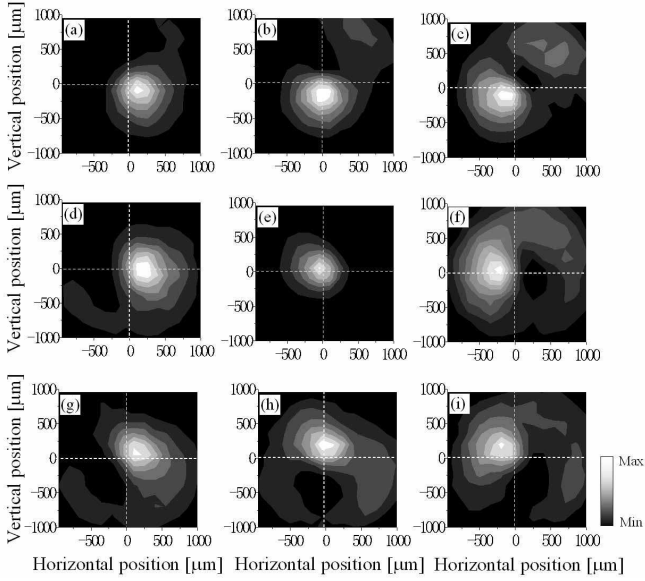}}\\
\end{center}
\caption{Maps of the coincidence count rates obtained by scanning hologram A with nine different positions of hologram B. In (e), the dislocation of hologram B is almost on the optical axis. The reference position of hologram B was taken to be $(r_{B},\theta_{B})=(0 \mu m,0)$. The positions of hologram B relative to this position are (a): $(200 \mu$m,$3\pi /4)$, (b): $(200 \mu$m,$\pi /2)$, (c): $(200 \mu$m,$\pi /4)$, (d): $(200 \mu$m,$\pi)$, (f): $(200 \mu$m,$0)$, (g):  $(200 \mu$m,$\pi /2)$, (h): $(200 \mu m, -\pi /2)$, (i): $(200 \mu m, -\pi /4)$.}
\label{fig:fig6}
\end{figure}
\begin{figure}[tbp]
\begin{center}
\scalebox{0.2}[0.2]{\includegraphics{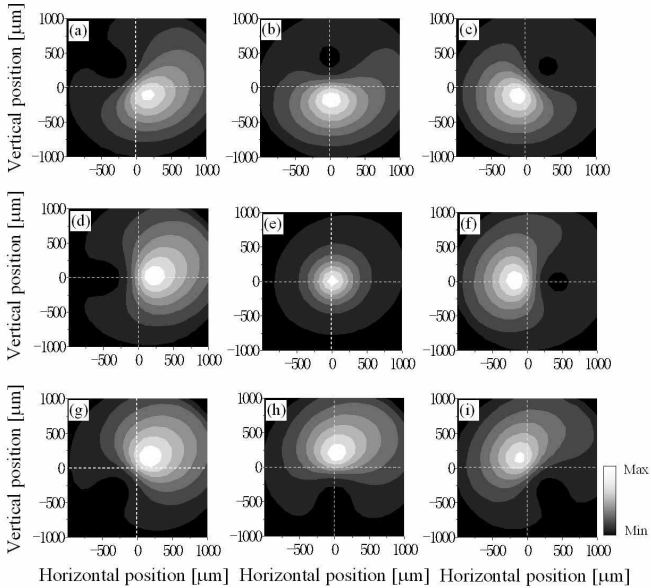}}\\
\end{center}
\caption{The simulated coincidence count probabilities for nine different positions of hologram B. The positions of hologram B $(r_{B},\theta_{B})$ are (a):  $(200 \mu$m, $3\pi /4)$, (b): $(200 \mu$m,$\pi /2)$, (c): $(200 \mu$m,$\pi /4)$, (d): $(200 \mu $m,$\pi)$, (e): $(0 \mu$m,$0)$, (f): $(200 \mu$m,$0)$, (g): $(200 \mu$m,$\pi /2)$, (h): $(200 \mu$m,$-\pi /2)$, (i): $(200 \mu$m,$-\pi /4)$. }
\label{fig:fig7}
\end{figure}
In order to determine the remaining parameters, we shifted the position of hologram B and measured the maps of coincidence count rates by scanning hologram A. The results for nine different positions of hologram B are shown in Fig. \ref{fig:fig6}. From all of these maps, Fig. \ref{fig:fig6}(e), which was measured by shifting hologram B by 200 $\mu$m upwards, exhibits a particularly symmetric pattern. For this condition, the minimum point $(\omega^{2}/(2r_{B}),\pi+\delta + \theta_{B})$ should be far from the center and thus $r_B$ should be close to zero. That is, the dislocation of hologram B is almost on the optical axis. As a result, for the original hologram B position in Fig. \ref{fig:fig5}, $r_B$ was confirmed to be 200 $\mu$m and $\theta_B$ was determined to be $-\pi/2$. We have already seen that $\theta_{B}+\delta = \pi /2$. Consequently, $\delta$ is calculated to be $\pi$.

The simulated coincidence count probabilities calculated following the same procedure as that of Fig. \ref{fig:fig5}(b) are shown in Fig. \ref{fig:fig7}. In the simulation, the maximum point appears on the opposite side of the origin from the minimum point, and both points move as the relative position of the hologram B changes. This is due to the non-classical correlations in the different non-orthogonal basis sets. These characteristics are observed in the experimental results (Fig. \ref{fig:fig6}). 

\section{VI. Verification of entanglement}

We also tried to verify the non-local correlation using the CHSH inequality\cite{Kwiat}, using the well-defined optical axes of the entangled modes. In the original discussion\cite{CHSH}, Clauser and co-authors assumed that two photons are sent along two opposite paths A and B, and the polarizations of those photons were analyzed using  polarization analyzer in both paths. Those polarization analyzers were characterized by a parameter $\theta$, and they discriminate the photons into two orthogonal channels, channel $\theta$ and $\overline{\theta}$, where the photons are detected. Under the `fair sampling assumption' \cite{Garuccio}, a correlation function $E(\theta_{A},\theta_{B})$ is defined as follows:
\begin{eqnarray} 
&&E(\theta_{A},\theta_{B})\nonumber\\
&=&\frac{C(\theta_{A},\theta_{B})+C(\overline{\theta_{A}},\overline{\theta_{B}})-C(\overline{\theta_{A}},\theta_{B})-C(\theta_{A},\overline{\theta_{B}})}
{C(\theta_{A},\theta_{B})+C(\overline{\theta_{A}},\overline{\theta_{B}})+C(\overline{\theta_{A}},\theta_{B})+C(\theta_{A},\overline{\theta_{B}})}\nonumber.\\
\end{eqnarray}

Here, $C(\theta_{A},\theta_{B})$ is the coincident count rate of photons in channel $\theta_{A}$ and  $\theta_{B}$. Then, the CHSH inequality,
\begin{equation} 
|S|=|E(\theta_{A},\theta_{B})-E(\theta'_{A},\theta_{B})+E(\theta_{A},\theta'_{B})+E(\theta'_{A},\theta'_{B})|\le 2,
\label{eqn:CHSH}
\end{equation} 
is valid for any local hidden variable theory.

In order to verify the violation of the CHSH inequality, we measured coincidence counts by changing
the relative azimuthal angle $\theta_A,\theta_B$ of the holograms A and B, while keeping the relative
distances constant, $r_A=r_B=r$. In this case, the coincidence probabilities are given as a function of $\theta_A$ and $\theta_B$ as follows,

\begin{equation} 
P(\theta_{A},\theta_{B})=\alpha^{2}\frac{4r^{4}+\omega^{4}+4r^{2}\omega^{2}\cos(\delta + \theta_{A}-\theta_{B})}{(2r^{2}+\omega^{2})^{2}}.
\end{equation}

In our experiment, we were just concerned with the two dimensional Hilbert space of each path A and B, since our experimental setup just measured the components of $|0\rangle$ and $|1\rangle$ in path A, and of $|0\rangle$ and $|-1\rangle$ in path B. In this case, we can substitute photon detection of the parameter $\theta^{\perp}=\theta + \pi$ for that in the channel $\overline{\theta}$. That is, the coincidence rate $C(\overline{\theta_{A}},\theta_B) $ in eq. (\ref{eqn:CHSH}) can be replaced by $C(\theta_{A}^{\perp},\theta_B) $\footnote{Here, we need to make an additional auxiliary assumption that the state from the source is independent of the analyzer settings\cite{Kwiat}. } .
\begin{eqnarray}
&&E(\theta_{A},\theta_{B})\nonumber\\
&=&\frac{C(\theta_{A},\theta_{B})+C(\theta_{A}^{\perp},\theta_{B}^{\perp})-C(\theta_{A}^{\perp},\theta_{B})-C(\theta_{A},\theta_{B}^{\perp})}
{C(\theta_{A},\theta_{B})+C(\theta_{A}^{\perp},\theta_{B}^{\perp})+C(\theta_{A}^{\perp},\theta_{B})+C(\theta_{A},\theta_{B}^{\perp})}.\nonumber\\
\label{eqn:CHSH2}
\end{eqnarray}
If it is assumed that the channel loss is independent of the angle $\theta$, then we can calculate
the correlation function $E(\theta_{A},\theta_{B})$ by replacing the coincidence rates $C(\theta_{A},\theta_{B})$ by $P(\theta_{A},\theta_{B})$ in eq. (\ref{eqn:CHSH2}).
When we set the parameters $\theta_{A}=-\pi/4$, $\theta'_{A}=\pi/4$, $\theta_{B}=-\pi/2$, $\theta'_{B}=0$ and $r=\omega/\sqrt{2}$, we obtain $|S|=2\sqrt{2} \sim 2.828 (>2)$

\begin{table}[tbp]
\begin{ruledtabular}
\begin{tabular}{ccccccc}
 $\theta_{A}$ & $\theta_{B}$ & $C(\theta_{A},\theta_{B}) $ & $C(\theta_{A},\theta_{B}^{\perp})$ & $C(\theta_{A}^{\perp},\theta_{B})$ & $C(\theta_{A}^{\perp},\theta_{B}^{\perp})$ \\
\hline
$-\pi /4$ & $-\pi /2 $ &53 &371 &449 &60 \\
$-\pi /4$ & $0$ & 120&256 &221 & 189 \\
$\pi /4$ & $-\pi /2$ & 261& 102& 203& 221 \\
$\pi /4$ 
& $0$ &4 &406 &428 &26  
\end{tabular}
\end{ruledtabular}
\caption{Coincidence count rates measured over 5 seconds.}
\end{table}
We performed experiments with the parameters $r=200 \mu$m, $\theta_{A}=-\pi/4$, $\theta'_{A}=\pi/4$, $\theta_{B}=-\pi/2, \theta'_{B}=0$. The obtained coincidence rates over 5 s accumulation period are shown in Table 1. Note that these data are measured independently from the data shown in Fig. \ref{fig:fig6}. The obtained $|S|$ was 2.127, which is larger than 2, indicating that the CHSH inequality has been violated
\footnote{$|S|$ was 2.134 when the raw coincidence rates in table 1 were compensated by single count rates.}.

The reason why $|S|$ was smaller than the maximum value 2.828 seems to be mainly because the value for $r$ we used, $r=200 \mu$m, was smaller than the optimum position of $r=283 \mu$m. In fact, $|S|$ is calculated to be 2.26 for $r=200\mu$m, which is close to the experimental result.

\section{VII. Conclusion}

In conclusion, we have proposed a method for determining the position of the hologram dislocations relative to the optical axes of entangled Laguerre Gaussian modes. In our method, the coincidence count rate distribution is obtained by scanning the position of one of the holograms, and then the relative position of the hologram dislocation is identified quantitatively from the positions of the minimum and maximum coincidence count rates. We tested this method experimentally with the entangled photons generated thorough the SPDC process and succeeded in determining the positions of the hologram dislocations relative to the optical axes of each mode. Finally, an experiment investigating the CHSH inequality was performed using the well-determined optical axes of the entangled modes, and the obtained value $|S|=2.127$, which is slightly smaller than the theoretically predicted value of 2.26, demonstrates violation of the CHSH inequality.

In the theory part, we neglected the radial index of the LG modes and derived eq. (5). It should be noted that the proposed method based on eq. (5) is useful as is shown in the experimental demonstration. However, the further studies on the effect of radial indexes and other `outer modes \cite{Molina-Terriza-PRL2004}' on the entanglement in LG modes as well as new experimental method to avoid those effects seem interesting and important. 

The authors would like to thank J. Arlt for providing us the reference \cite{Arlt-PhDthesis},  H. F. Hofmann, K. Tsujino, R. Okamoto for their discussion, M. Moriya for gold coating, H. Ohminato, T. Yonemura for their technical support. This work was supported in part by 
Core Research for Evolutional Science and Technology, Japan Science and Technology Agency,
Grant-in-Aid of Japan Science Promotion Society ( No.17684021 and No.14750028 ) and the 21st century COE program.

\end{document}